\documentstyle[epsfig,12pt]{article}

\def\lord{$ \raisebox{-.3ex}{$\stackrel{<}{_{\sim}}$} $}

\title{
{\bf PRIMORDIAL BLACK HOLES AND HOT MATTER}}
\author{{JOSEPH I. KAPUSTA} \vspace*{0.1in}\\
 {\it School of Physics and Astronomy, University of Minnesota}\\
 {\it Minneapolis, MN 55455, USA}}

\date{}

\begin{document}

\maketitle

\begin{abstract}

Microscopic black holes explode with their temperature varying inversely as
their mass.  Such explosions would lead to the highest temperatures in the
present universe, all the way to the Planck energy.  The possibility
that a quasi-stationary shell of hot matter surrounds these black holes
has recently been proposed and studied with relativistic Boltzmann
transport equations and with relativistic viscous fluid dynamics.  For
example, a black hole with a mass of 10$^{10}$ g has a Hawking
temperature of 1 TeV, a Schwarszchild radius of 1.6$\times$10$^{-5}$
fm, a luminosity of 7$\times$10$^{27}$ erg/s, and has less than 8
minutes to live.  It is an outstanding theoretical challenge to
describe the conditions exterior to such microscopic black holes and a
great challenge to finally detect them in the new millennium.

\end{abstract}

Hawking radiation from black holes \cite{Hawk} is of fundamental interest
because it relies on the application of relativistic quantum field theory in the
presence of the strong field limit of gravity, a so far unique situation.  It is
also of great interest because of the temperatures involved.  A black hole with
mass $M$ radiates thermally with a Hawking temperature
\begin{equation}
T_H = \frac{m_{\rm P}^2}{8\pi M}
\end{equation}
where $m_{\rm P} = G^{-1/2} = 1.22\times 10^{19}$ GeV is the Planck mass.
(Units are $\hbar = c = k_{\rm B} = 1$.)  In order for the black hole
to evaporate it must have a temperature greater than that of the present-day
black-body radiation of the universe of 2.7 K = 2.3$\times 10^{-4}$ eV.
This implies that $M$ must be less than $1\%$ of the mass of the Earth,
hence the black hole most likely would have been formed primordially and not
from stellar collapse.  The black hole temperature eventually goes to infinity
as its mass goes to zero, although once $T_H$ becomes comparable to the
Planck mass the semi-classical calculation breaks down and the regime of full
quantum gravity is entered.  Only in two other situations are such enormous
temperatures achievable: in the early universe ($T$ similarly asymptotically
high) and in central collisions of heavy nuclei like gold or lead ($T = 500$ MeV
is expected at the RHIC (Relativistic Heavy Ion Collider) just completed at
Brookhaven National Laboratory and $T = 1$ GeV is expected at the LHC (Large
Hadron Collider) at CERN to be completed in 2005).  Supernovae and newly formed
neutron stars are unlikely to ever exceed a temperature of 50 MeV.
To set the scale from fundamental physics, we note that the spontaneously broken
chiral symmetry of QCD gets restored in a phase transition/rapid crossover at a
temperature around 160 MeV, while the spontaneously broken gauge symmetry in the
electroweak sector of the standard model gets restored in a phase
transition/rapid crossover at a temperature around 100 GeV.  The fact that
temperatures of the latter order of magnitude will never be achieved in a
terrestrial experiment should motivate us to study the fate of primordial black
holes during the final minutes of their lives when their temperatures have risen
to 100 GeV and above.  The fact that primordial black holes have not yet been
observed \cite{review} should not be viewed as a deterrent, but rather as a
challenge!

There are at least two intuitive ways to think about Hawking radiation from
black holes.  One way is vacuum polarization.  Particle-antiparticle pairs are
continually popping in and out of the vacuum, usually with no observable effect.
In the presence of matter, however, their effects can be observed.  This is the
origin of the Lamb effect first measured in atomic hydrogen in 1947.  When pairs
pop out of the vacuum near the event horizon of a black hole one of them may be
captured by the black hole and the other by necessity of conservation laws will
escape to infinity with positive energy.  The black hole therefore has lost
energy - it radiates.  Due to the general principles of thermodynamics applied
to black holes it is quite natural that it should radiate thermally.  An
intuitive argument that is more quantitative is based on the uncertainty
principle.  Suppose that we wish to confine a massless particle to the vicinity
of a black hole.  Given that the average momentum of a massless particle at
temperature $T$ is approximately $\pi T$, the uncertainty principle requires
that confinement to a region the size of the Schwarzschild diameter places a
restriction on the minimum value of the temperature.
\begin{equation}
\pi T \cdot 2 r_S \geq 1/2
\end{equation}
The minimum is actually attained for the Hawking temperature.  The various
physical quantities are related as $r_S = 2M/m_{\rm P}^2 =
1/4\pi T_H$.

The number of particles of spin $s$ emitted with energy $E$ per unit time is
given by the formula
\begin{equation}
\frac{dN_s}{dEdt} = \frac{\Gamma_s}{2\pi} \,
\frac{1}{\exp(E/T_H)-(-1)^{2s}} \, .
\end{equation}
All the computational effort really goes into calculating the absorption
coefficient $\Gamma_s$ from a relativistic wave equation in the presence of a
black hole.  Integrating over all particle species yields the luminosity.
\begin{equation}
L = -\frac{dM}{dt} = \alpha(M) \frac{m_{\rm P}^4}{M^2} =
64 \pi^2 \alpha(T_H) T_H^2 \, .
\end{equation}
Here $\alpha(M)$ is a function reflecting the species of particles available
for creation in the gravitational field of the black hole.  It is generally
sufficient to consider only those particles with mass less than $T_H$;
more massive particles are exponentially suppressed by the Boltzmann factor.
Then
\begin{equation}
\alpha = 2.011\times 10^{-8} \left[ 4200 N_0 + 2035 N_{1/2} + 835 N_1 + 95 N_2
\right]
\end{equation}
where $N_s$ is the net number of polarization degrees of freedom for all
particles with spin $s$ and with mass less than $T_H$.  The coefficients for
spin 1/2, 1 and 2 were computed by Page \cite{Page} and for spin 0 by Sanchez
\cite{Sanchez}.  In the standard model $N_0 = 4$ (Higgs), $N_{1/2} = 90$ (three
generations of quarks and leptons), $N_1 = 24$ (SU(3)$\times$SU(2)$\times$U(1)
gauge theory), and $N_2 = 2$ (gravitons).  This assumes $T_H$ is greater than
the temperature for the electroweak gauge symmetry restoration.
Numerically $\alpha(T_H > 100 \,{\rm GeV}) = 4.43\times 10^{-3}$.  Starting with
a black hole of temperature $T_H$, the time it takes to evaporate/explode is
\begin{equation}
\Delta t = \frac{m_{\rm P}^2}{3 \alpha(T_H) (8\pi T_H)^3} \, .
\end{equation}
This is also the characteristic time scale for the rate of change of the
luminosity of a black hole with temperature $T_H$.

At present a black hole will explode if $T_H > 2.7$ K.  This corresponds
to a critical mass of $M < 4.6\times 10^{25}$ g which is approximately
1\% of the mass of the Earth.
More massive black holes are cooler and therefore will absorb more matter and
radiation than they radiate, hence grow with time.  Taking into account emission
of gravitons, photons, and neutrinos a critical mass black hole today has a
Schwarszchild radius of 68 microns and a lifetime of $2\times10^{43}$ years.

There is some uncertainty over whether the particles scatter from each
other after being emitted, perhaps even enough to allow a fluid description of
the wind coming from the black hole.  Let us examine what might happen as the
black hole mass decreases and the associated Hawking temperature increases.

When $T_H \ll m_{\rm e}$ (electron mass) only photons, gravitons,
and neutrinos will be created with any significant probability.  These
particles will not interact with each other but will be emitted into
the surrounding space with the speed of light \cite{Hawk,Page}.
Even when $T_H \approx m_{\rm e}$ the Thomson cross section is too
small to allow the photons to scatter very frequently in the rarified
electron-positron plasma around the black hole.  This may change when
$T_H \approx 100$ MeV when muons and charged pions are created in
abundance.  At somewhat higher temperatures hadrons are copiously produced and
local thermal equilibrium may be achieved, although exactly how is an unsettled
issue.  Are hadrons emitted directly by the black hole?  If so, they will be
quite abundant at temperatures of order 150 MeV because their mass spectrum
rises exponentially (Hagedorn growth as seen in the Particle Data Tables
\cite{PDG}).  Because they are so massive they move nonrelativistically and may
form a very dense equilibrated gas around the black hole.  But hadrons are
composites of quarks and gluons, so perhaps quarks and gluon jets are emitted
instead?  These jets must decay into the observable hadrons on a typical length
scale of 1 fm and a typical time scale of 1 fm/c.  This was first studied by
MacGibbon and Webber \cite{MW} and MacGibbon and Carr \cite{MC}.  Subsequently
Heckler \cite{Ha} argued that since the emitted quarks and gluons are so densely
packed outside the event horizon that they are not actually fragmenting into
hadrons in vacuum but in something more like a quark-gluon plasma, so perhaps
they thermalize.  He also argued that QED bremsstrahlung and pair production
were sufficient to lead to a thermalized QED plasma when $T_H$ exceeded 45 GeV
\cite{Hb}.  These results are somewhat controversial and need to be confirmed.
The issue really is how to describe the emission of wavepackets via the Hawking
mechanism when the emitted particles are (potentially) close enough to be
mutually interacting.  A more quantitative treatment of the particle
interactions on a semiclassical level was carried out by Cline, Mostoslavsky and
Servant \cite{cline}.  They solved the relativistic Boltzmann equation with QCD
and QED interactions in the relaxation-time approximation. It was found that
significant particle scattering would lead to a photosphere though not perfect
fluid flow.

The picture proposed by Heckler is as follows:  Hawking radiation
is emitted as usual in the vicinity of the Schwarzschild radius.  Bremsstrahlung
scattering leads to approximate thermalization and fluid flow at a radius
$r_0 > r_S$.  Thermal contact is lost at a photosphere radius of
$r_p > r_0$.  By then most electrons and positrons have annihilated; the energy
goes off to infinity in the form of photons with average energy much less than
$T_H$.

Figures 1-3 show some results from Cline et al.  The number of scatterings in the
QCD shell of matter interior to the hadronization radius grows significantly
with increasing $T_H$.  Indeed the total increase in particle production due to
QCD bremsstrahlung and to quark and gluon fragmentation goes linearly with
$T_H$, while the outer radius of the QCD shell of matter grows logarithmically.
\begin{figure}
\centerline{\epsfig{figure=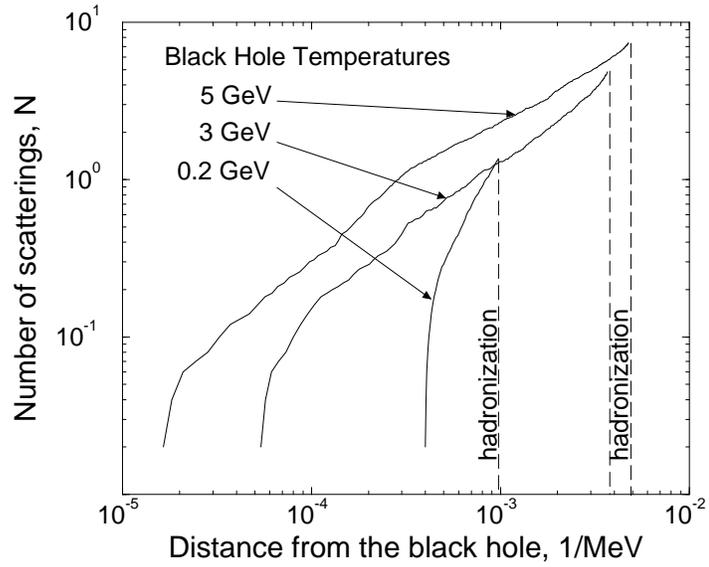,width=10.0cm}}
\caption{Average number of scatterings in QCD photosphere as a function
of radius for several black hole temperatures.  From \cite{cline}.}
\end{figure}
\begin{figure}
\centerline{\epsfig{figure=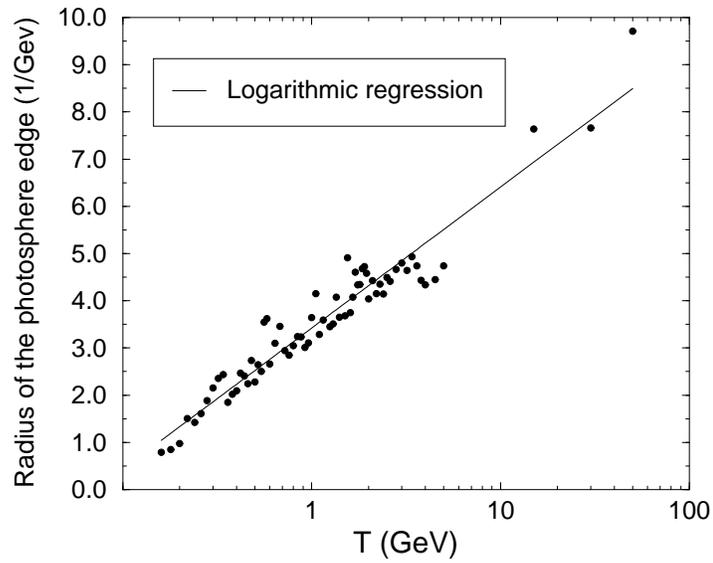,width=10.0cm}}
\caption{Radius of the outer surface of the QCD photosphere versus
logarithm of the black hole temperature.  From \cite{cline}.}
\end{figure}
\begin{figure}
\centerline{\epsfig{figure=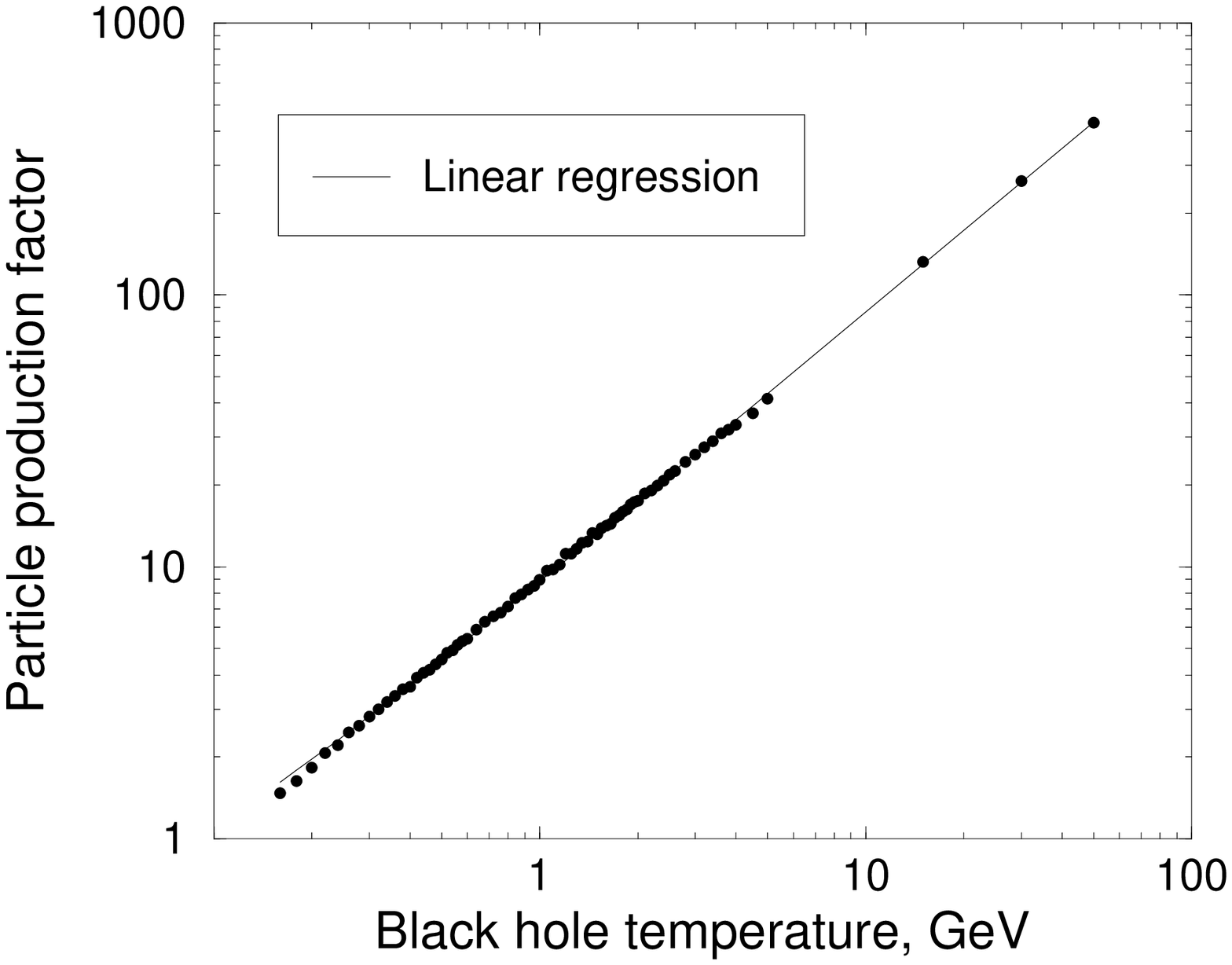,width=10.0cm}}
\caption{Total particle production in the QCD photosphere versus
black hole temperature.  From \cite{cline}.}
\end{figure}

Rather than pursuing the Boltzmann transport equation I will now attempt a 
description in terms of relativistic viscous fluid dynamics \cite{me}.
Let us assume that a primordial black hole is surrounded by a shell of
expanding matter in approximately local thermal equilibrium when $T_H$ is large
enough.  A detailed description of how this situation comes to be is a difficult
problem as discussed above and needs further investigation.  The
applicability can and will be checked {\em a posteriori}.  The relativistic
imperfect fluid equations describing a steady-state, spherically symmetric flow
with no net baryon number or electric charge and neglecting gravity
(see below) are $T^{\mu\nu}_{\;\;\;\;;\nu} =$ {\em black hole source}.  The
nonvanishing components of the energy-momentum tensor in radial coordinates are
\cite{MTW}
\begin{eqnarray}
T^{00}&=& \gamma^2 (P+\epsilon) -P + v^2 \Delta T_{\rm diss} \nonumber \\
T^{0r}&=& v\gamma^2 (P+\epsilon) + v \Delta T_{\rm diss} \nonumber \\
T^{rr}&=& v^2\gamma^2 (P+\epsilon) +P + \Delta T_{\rm diss}
\end{eqnarray}
representing energy density, radial energy flux, and radial momentum flux,
respectively, in the rest frame of the black hole.  Here
$v$ is the radial velocity with $\gamma$ the corresponding Lorentz factor, $u =
v\gamma$, $\epsilon$ and $P$ are the local energy density and pressure, and
\begin{equation}
\Delta T_{\rm diss} = -\frac{4}{3}\eta \gamma^2 \left( \frac{du}{dr}
-\frac{u}{r}\right) - \zeta \gamma^2 \left( \frac{du}{dr}
+\frac{2u}{r}\right) \, ,
\end{equation}
where $\eta$ is the shear viscosity and $\zeta$ is the bulk viscosity.  A
thermodynamic identity gives $Ts = P + \epsilon$ for zero chemical potentials,
where $T$ is temperature and $s$ is entropy density.  There are two independent
differential equations of motion to solve for the functions $T(r)$ and $v(r)$.
These may succinctly be written as
\begin{eqnarray}
\frac{d}{dr} \left( r^2 T^{0r} \right) &=& 0 \nonumber \\
\frac{d}{dr} \left( r^2 T^{rr} \right) &=& 0 \, .
\end{eqnarray}

An integral form of these equations is probably more useful since it can readily
incorporate the input luminosity $L_i$ from the black hole.  The first
represents the equality of the energy flux passing through a sphere of radius r
with the luminosity of the black hole.
\begin{equation}
4\pi r^2 T^{0r} = L_i
\end{equation}
The second follows from integrating a linear combination of the differential
equations.  It represents the combined effects of the entropy from the black
hole together with the increase of entropy due to viscosity.
\begin{equation}
4\pi r^2 u s  = 4\pi \int_{r_i}^r dr' \, r'^2 \frac{1}{T}\left[
\frac{8}{9} \eta \left( \frac{du}{dr'} - \frac{u}{r'} \right)^2
+ \zeta \left( \frac{du}{dr'} + \frac{2u}{r'} \right)^2 \right]
+ \frac{L_i}{T_H}
\end{equation}
The term $L_i/T_H$ arises from equating the entropy per unit time lost by the
black hole $-d S_{\rm bh}/dt$ with that flowing into the matter.  Using the area
formula for the entropy of a black hole, $S_{\rm bh} = m_{\rm P}^2 \pi r_S^2 =
4\pi M^2/m_{\rm P}^2$, and identifying $-dM/dt$ with the luminosity, the entropy
input from the black hole is obtained.
The above pair of equations are to be applied beginning at some radius $r_i$
greater than the Schwarzschild radius $r_S$, that is, outside the quantum
particle production region of the black hole.
The radius $r_i$ at which the imperfect fluid equations are first applied should
be chosen to be greater than the Schwarzschild radius, otherwise the computation
of particle creation by the black hole would be invalid.  It should not be too
much greater, otherwise particle collisions would create more entropy than is
accounted for by the equation above.  The energy and entropy flux into the fluid 
come from quantum particle creation by the black hole at temperature $T_H$. 
Gravitational effects are of order $r_S/r$, hence
negligible for $r > (5-10)r_S$.

Determination of the equation of state as well as the two viscosities for
temperatures ranging from MeV to TeV and more is a formidable task.  Here we
shall consider two interesting limits and then a semi-realistic situation.  A
realistic, quantitative description of the relativistic black hole wind,
including the asymptotic observed particle spectra, is currently under 
investigation by my graduate student Ramin Daghigh and myself.

First, consider the nonviscous limit (like milk) with an equation of state
$\epsilon = aT^4$, $s = (4/3)aT^3$, and $\eta = \zeta = 0$.  This is equivalent
to assuming that the mean free paths of the particles are all small compared to
the length scale over which the temperature and other thermodynamic quantities
change significantly.  A scaling solution, valid when $\gamma \gg 1$, is $T(r) =
T_0(r_0/r)$ and $\gamma(r) = \gamma_0(r/r_0)$, where $\gamma_0 T_0 = T_H$.  The
$r_0$ is any reference radius satisfying the stated criterion.

Second, consider the highly viscous isoergic limit (like honey) in the sense
that the flow velocity approaches a limiting value $v_0$ at large $r$.  This
requires a power-like equation of state $\epsilon \propto T^{\delta}$ and
viscosities $\eta \propto \zeta \propto T^{\delta/2}$.  It results in the
scaling solution $T(r) = T_0 (r_0/r)^{2/\delta}$.  This is not very realistic: a
massless gas with dimensionless coupling constants and $\delta = 4$ would
require viscosities of order $T^2$ whereas one would expect $T^3$ on dimensional
grounds.

Now consider a semi-realistic situation with $\epsilon = aT^4$, $s = (4/3)aT^3$,
$\eta = b_ST^3$, and $\zeta = b_BT^3$.  This is typical of relativistic gases 
with dimensionless coupling constants, although quantum effects will give
logarithmic corrections \cite{baym,tau}.
A scaling solution, valid at large radii when $\gamma \gg 1$, is
$T(r) = T_0 (r_0/r)^{2/3}$ and $\gamma(r) = \gamma_0 (r/r_0)^{1/3}$.
This $r$-dependence of $T$ and $\gamma$ is exactly what was conjectured by 
Heckler \cite{Hb}.

Is the semi-realistic situation described above really possible?  Can
approximate local thermal equilibrium, if once achieved, be maintained?
The requirement is that the inverse of the local volume expansion rate
$\theta = u^{\mu}_{\;\; ;\mu}$ be
comparable to or greater than the relaxation time for thermal equilibrium
\cite{MTW}.  Expressed in terms of a local volume element $V$ and proper time
$\tau$ it is $\theta = (1/V)dV/d\tau$, whereas in the rest frame of the black
hole the same quantity can be expressed as $(1/r^2)d(r^2 u)/dr$.  Explicitly
\begin{equation}
\theta = \frac{7\gamma_0}{3r_0}\left(\frac{r_0}{r}\right)^{2/3}
= \frac{7\gamma_0}{3r_0T_0} T \, .
\end{equation}
Of prime importance in achieving and maintaining local thermal equilibrium in a
relativistic plasma are multi-body processes such as $2 \rightarrow 3$ and
$3 \rightarrow 2$, etc.  This has been well-known when calculating quark-gluon
plasma formation and evolution in high energy heavy ion collisions \cite{klaus}
and has been emphasized in ref. \cite{Ha,Hb} in the context of black hole
evaporation.  This is a formidable task in the standard model with its 16
species of particles.  Instead we make three estimates for the requirement that
local thermal equilibrium be maintained. The first and simplest estimate is to
require that the thermal DeBroglie wavelength of a massless particle,
$1/3T$, be less than $1/\theta$.   The second estimate is to require that
the Debye screening length for each of the gauge groups in the standard model be
less than $1/\theta$. The Debye screening length is the inverse of the Debye
screening mass $m^{\rm D}_n$ where $n =1, 2, 3$ for the gauge groups U(1),
SU(2), SU(3).  Generically $m^{\rm D}_n \propto g_nT$ where $g_n$ is the gauge
coupling constant and the coefficient of proportionality is essentially the
square root of the number of charge carriers \cite{kapbook}.  For example, for
color SU(3) $m^{\rm D}_3 = g_3 \sqrt{1+N_{\rm f}/6}\,T$
where $N_{\rm f}$ is the number of light quark flavors at the temperature $T$.
The numerical values of the gauge couplings are: $g_1 = 0.344$, $g_2 = 0.637$,
and $g_3 = 1.18$ (evaluated at the scale $m_Z$) \cite{PDG}.  So within a factor
of about 2 we have $m^{\rm D} \approx T$. The third and most relevant estimate
is the mean time between two-body collisions in the standard model for
temperatures greater than the electroweak symmetry restoration temperature.
This mean time was calculated by Carrington and me \cite{tau} in the process of
calculating the viscosity in the relaxation time approximation.  Averaged over
all particle species in the standard model one may infer from that paper an
average time of $3.7/T$.  Taking into account multi-body reactions would
decrease that by about a factor of two to four.  All three of these estimates
are consistent within a factor of 2 or 3.  The conclusion to be drawn is that
local thermal equilibrium should be achieved when
$\theta \lord T$. Once thermal equilibrium is achieved it is not lost because 
$\theta/T$ is independent of $r$.  The picture that emerges is that of an 
imperfect fluid just marginally kept in local equilibrium by viscous forces.

The hot shell of matter surrounding a primordial black hole provides a
theoretical testing ground rivaled only by the big bang itself.  In addition to
the questions already raised, one may contemplate baryon number violation at
high temperature and how physics beyond the standard model might be
important in the last few minutes in the life of a primordial black hole.
To illustrate this I have plotted a semi-realistic parametrization of the 
equation of state in figure 4.
\begin{figure}
\centerline{\epsfig{figure=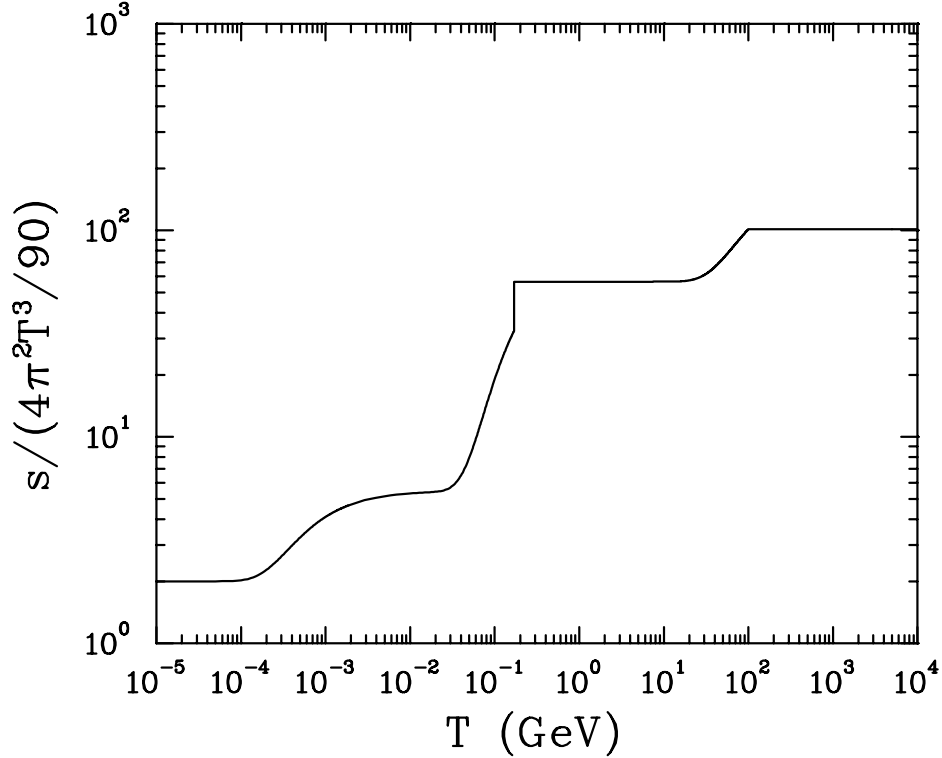,width=10.0cm,angle=90}}
\caption{Entropy density as a function of temperature, excluding neutrinos
and gravitons.  It is assumed that the QCD phase transition is first order
and the EW phase transition is second order.}
\end{figure}
Gravitons and neutrinos are not included.  I 
have assumed a second order electroweak phase transition at a temperature of 100 
GeV.  Above that temperature the standard model has 101.5 effective massless 
bosonic degrees of freedom (as usual fermions count as 7/8 of a boson).  There 
is a first order QCD phase transition at a temperature of 160 MeV.  The number 
of effective massless bosonic degrees of freedom changes from 47.5 just above 
this critical temperature (u, d, s quarks and gluons) to 7.5 just below it 
(representing the effects of all the massive hadrons in the particle data 
tables) \cite{OK}.  Below 30 MeV only electrons, positrons, and photons remain, 
and finally below a few hundred keV only photons survive in any appreciable 
number.  This is a rather crude description of the real equation of state that 
is still a matter of theoretical uncertainty.  One needs knowledge of $s(T), 
\eta(T), \zeta(T)$ over a huge range of $T$!  Of course, these are some of the 
quantities one hopes to obtain experimental information on from observations of 
exploding black holes.

The parallelism with high energy nucleus-nucleus collisions cannot go unnoticed 
and without a certain amount of cross fertilization.  In heavy ion physics the 
best theories are founded on relativistic fluid dynamics and on microscopic 
transport models \cite{qm}.  Both approaches can be fruitfully employed to 
describe the conditions outside the event horizon of a microscopic black hole.  
There is much work to be done in this area.  Finally, such black holes may 
contribute to the highest energy cosmic rays whose origin is a long-standing 
puzzle.  Experimental discovery of exploding black holes will be one of the 
great challenges in the new millennium!

\section*{Acknowledgements}

This work was supported by the US Department of Energy under grant
DE-FG02-87ER40328.


\begin{thebibliography} {99}

\bibitem{Hawk} S. W. Hawking, Nature (London) {\bf 248}, 30 (1974); Commun.
Math. Phys. {\bf 43}, 199 (1975).

\bibitem{review} B. J. Carr and J. H. MacGibbon, Phys. Rep. {\bf 307}, 141
(1998).

\bibitem{Page} D. N. Page, Phys. Rev. D {\bf 13}, 198 (1976).

\bibitem{Sanchez} N. Sanchez, Phys. Rev. D {\bf 18}, 1030 (1978); the
coefficient was extracted from figure 5 of this paper to within about 5\%
accuracy.

\bibitem{PDG} Particle Data Group: {\it Review of Particle Physics}, Eur. Phys.
J. {\bf C3}, 1 (1998).

\bibitem{MW} J. H. MacGibbon and B. R. Webber, Phys. Rev. D {\bf 41}, 3052 (1990).

\bibitem{MC}  J. H. MacGibbon and B. J. Carr, Astrophys. J. {\bf 371}, 447 (1991).

\bibitem{Ha} A. F. Heckler, Phys. Rev. Lett. {\bf 78}, 3430 (1997).

\bibitem{Hb} A. F. Heckler, Phys. Rev. D {\bf 55}, 480 (1997).

\bibitem{cline} J. Cline, M. Mostoslavsky, and G. Servant, Phys. Rev. D
{\bf 59}, 063009 (1999).

\bibitem{me} J. I. Kapusta, Phys. Rev. Lett., in press (preprint
astro-ph/0008222).

\bibitem{MTW} C. W. Misner, K. S. Thorne, and J. A. Wheeler, {\em Gravitation}
(W. H. Freeman and Company, San Francisco, 1973).

\bibitem{baym} G. Baym, H. Monien, C. J. Pethick, and D. G. Ravenhall,
Phys. Rev. Lett. {\bf 64}, 1867 (1990).

\bibitem{tau}  M. E. Carrington and J. I. Kapusta, Phys. Rev. D {\bf 47}, 5304
(1993).

\bibitem{klaus} K. Kinder-Geiger, Phys. Rep. {\bf 258}, 237 (1995);
S.M.H. Wong, Phys. Rev. D {\bf 54}, 2588 (1996).

\bibitem{kapbook} J. I. Kapusta, {\em Finite Temperature Field Theory}
(Cambridge University Press, Cambridge, England, 1989).

\bibitem{OK} J. I. Kapusta and K. A. Olive, Phys. Lett. {\bf B209},
295 (1988).

\bibitem{qm} See, for example, the proceedings of the Quark Matter series of
international conferences, the most recent in print being: {\it Proceedings of
Quark Matter `99}, Nucl. Phys. {\bf A661}, (1999).

\end{thebibliography}
\end{document}